\documentclass{article}
\usepackage{orcidlink}
\usepackage[margin=4cm]{geometry}
\usepackage{algpseudocode}
\usepackage{algorithm}
\usepackage{arxiv}
\usepackage[utf8]{inputenc} 
\usepackage[T1]{fontenc}    
\usepackage{hyperref}       
\usepackage{url}            
\usepackage{graphicx}
\usepackage{plantuml}
\usepackage{multirow}
\usepackage{booktabs}       
\usepackage{flafter}		
\usepackage{longtable}
\usepackage{acronym}
\usepackage[acronym]{glossaries}
\usepackage{diagbox}

\graphicspath{ {./images/} }
\title{Mitigating Hallucination | ZeroG: An Advanced Knowledge Management Engine}

\date{8 November 2024} 					

\author{ 
    \href{https://orcid.org/0000-0002-9064-3362}
    {\includegraphics[scale=0.06]{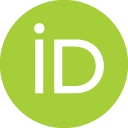}\hspace{1mm}Anantha Sharma$^*$}\\
	The A-Team (Charlotte, USA)\\
    \texttt{Anantha.Sharma@synechron.com}
    \And
    \href{https://orcid.org/0009-0003-8384-7521}
    {\includegraphics[scale=0.06]{orcid}\hspace{1mm}Sheeba Elizabeth John$^*$}\\
    The A-Team (Bengaluru, India)\\
    \texttt{Sheeba.John@synechron.com}
    \And
    \href{https://orcid.org/0009-0002-4952-7762}
    {\includegraphics[scale=0.06]{orcid}\hspace{1mm}Fatemeh Rezapoor Nikroo$^*$}\\
	 The A-Team (Montreal, Canada)\\
    \texttt{Fatemeh.Nikroo@synechron.com}\\
    \And
    \href{https://orcid.org/0009-0005-0948-7972}
    {\includegraphics[scale=0.06]{orcid}\hspace{1mm}Krupali Bhatt$^*$}\\
	 The A-Team (Montreal, Canada)\\
    \texttt{Krupali.Bhatt@synechron.com}\\
    \And
    \href{https://orcid.org/0009-0008-0286-9718}
    {\includegraphics[scale=0.06]{orcid}\hspace{1mm}Mrunal Zambre}\\
	 The A-Team (Intern, 2023)\\
\texttt{mrunaldeepak.zambre@sjsu.edu}\\
    \And
    \href{https://orcid.org/0009-0006-3374-6599}
    {\includegraphics[scale=0.06]{orcid}\hspace{1mm}Aditi Wikhe}\\
	 The A-Team (Intern, 2023)\\
    \texttt{aditiw2@illinois.edu}\\
}
     \renewcommand{\headeright}{}
 \renewcommand{\undertitle}{Synechron AI Practice}

\begin{document}
\maketitle
\def\thefootnote{*}\footnotetext{These authors contributed equally to this work.}

\begin{abstract}
The growth of digital documents presents significant challenges in efficient management and knowledge extraction. Traditional methods often struggle with complex documents, leading to issues such as hallucinations and high latency in responses from Large Language Models (LLMs). ZeroG, an innovative approach, significantly mitigates these challenges by leveraging knowledge distillation and prompt tuning to enhance model performance.

ZeroG utilizes a smaller model that replicates the behavior of a larger teacher model, ensuring contextually relevant and grounded responses, by employing a black-box distillation approach, it creates a distilled dataset without relying on intermediate features, optimizing computational efficiency. This method significantly enhances accuracy and reduces response times, providing a balanced solution for modern document management.

Incorporating advanced techniques for document ingestion and metadata utilization, ZeroG improves the accuracy of question-and-answer systems. The integration of graph databases and robust metadata management further streamlines information retrieval, allowing for precise and context-aware responses. By transforming how organizations interact with complex data, ZeroG enhances productivity and user experience, offering a scalable solution for the growing demands of digital document management.
\end{abstract}

\section{Introduction}
ZeroG significantly improves the quality of responses by a large margin by mitigating hallucinations through the implementation of knowledge distillation and prompt tuning, ensuring responses are accurate and grounded. We differ from \cite{Gou_2021}, which involves fine-tuning the student model, by utilizing a black-box distillation approach without fine-tuning. This approach reduces response latency and enhances overall system reliability. ZeroG leverages LLMs to generate Question and Answer (QnA) pairs from existing documents, storing them in a vector store. When a user query is received, similarity searches using techniques like MMR determine whether it can be addressed by pre-existing QnA pairs or requires a more tailored response using document-specific information.

Despite advances in natural language processing, traditional methods often struggle with real-time data updates and accurately handling complex documents, which include sensitive data. This paper explores transforming these presentations into markdown files for easier ingestion into vector stores, enhancing QnA accuracy without frequent reengineering. We also investigate integrating graph databases and utilizing document metadata to refine search and organization capabilities. By pre-generating question sets and caching commonly asked queries, the system streamlines responses, ensuring they are precise and contextually relevant. This paper presents some of the techniques we explored and employed to overcome current limitations in document and knowledge management, significantly improving productivity and effectiveness in handling complex document types.

\section{Methodology}
Our solution primarily focuses on utilizing out-of-the-box LLMs without any fine-tuning to extract knowledge from PowerPoint files, including presentations for RFP (Request for Proposal) responses and internal organizational case studies. This specialization enhances the system's ability to handle complex layouts and sensitive content effectively. 
Our methodology involves several key steps: creating a knowledge store, integrating flow orchestration with LangChain, and employing local models for enhanced privacy and performance. This setup allows the system to efficiently handle a wide range of queries, from simple factual questions to complex analytical tasks, by leveraging the combined strengths of the knowledge store and language models.

\subsection{Data}
The data encompasses internal organizational content, including RFP responses, case studies, and sales pitches in the form of presentation files. These documents often contain complex and sensitive information \cite{sharma2024tableguardsecuringstructured} that is crucial for business operations. Our objective is to empower end users within the organization to interact seamlessly with these documents while maintaining strict access controls. By implementing tailored access rights, we ensure that users receive secure and personalized access to the valuable insights these presentations offer, enhancing both efficiency and confidentiality in data handling.

\subsection{Document Preprocessing and Semantic Management}
To prepare sensitive documents for efficient integration, we convert PowerPoint presentations (PPTX/PPT) into Markdown format, ensuring compatibility and preserving content structure. Given the complex and inconsistent nature of these documents, parsing them into Markdown files allows us to capture the appropriate format and layout. Using our custom homegrown solution for reading PPT files and converting them to Markdown format, we ensure that the content is represented accurately to a large extent.

This preprocessing step performs the heavy lifting before the data is sent to the LLM for QnA generation, as the Markdown content is already structured and ready for processing. This conversion focuses on specific shape types, excluding images, and experiments with slide layouts to maintain readability. We utilize a graph database, such as Neo4j, to manage a master list of keywords and synonyms, enabling efficient storage and retrieval. Users can add new keywords, allowing the system to adapt to evolving needs.

During preprocessing, metadata like creation date and author information is extracted for improved document organization. We perform keyword matching against the master list, tagging content with relevant keywords and appending synonyms as hashtags, enhancing semantic relationships. Our tagging system is compatible with tools like Foam (an extension in VS Code) and Obsidian, linking related content to improve retrieval. Additionally, we capture document ownership details, including author name and contact information, to enhance accountability.

\subsection{Access Control and Data Privacy}
Access controls are crucial for restricting unauthorized access to sensitive information. In ZeroG, contextual mapping is employed to ensure users only receive responses aligned with their access rights. While all information resides in the vector database, it is filtered to prevent unauthorized access to sensitive commercial data.

The system maps the user’s query against their permissions to determine if they are authorized to access specific content. This filtering mechanism ensures that only users with the appropriate access can view sensitive data, maintaining privacy and confidentiality. By implementing such robust access controls, ZeroG effectively safeguards sensitive information while enabling efficient document management.

\subsection{Knowledge Store - Version Control and Review System}
To ensure effective document version control and collaboration, markdown files are integrated into a knowledge store that operates as a Git repository. This system facilitates efficient tracking of changes and enables multiple users to collaborate by adding, modifying, or deleting documents as necessary. After modifications, documents are subjected to a rigorous review process. Designated reviewers, who are subject matter experts, assess the changes for relevance, accuracy, and adherence to company standards. These reviewers have the authority to accept or reject modifications. Once approved, documents are transitioned to a read-only repository, ensuring that users access only validated and authoritative content, maintaining the integrity and reliability of the information.

\subsection{Document Ingestion}

\begin{figure}[h]
   \begin{minipage}{\textwidth}
     \centering
     \includegraphics[height=8.5cm]{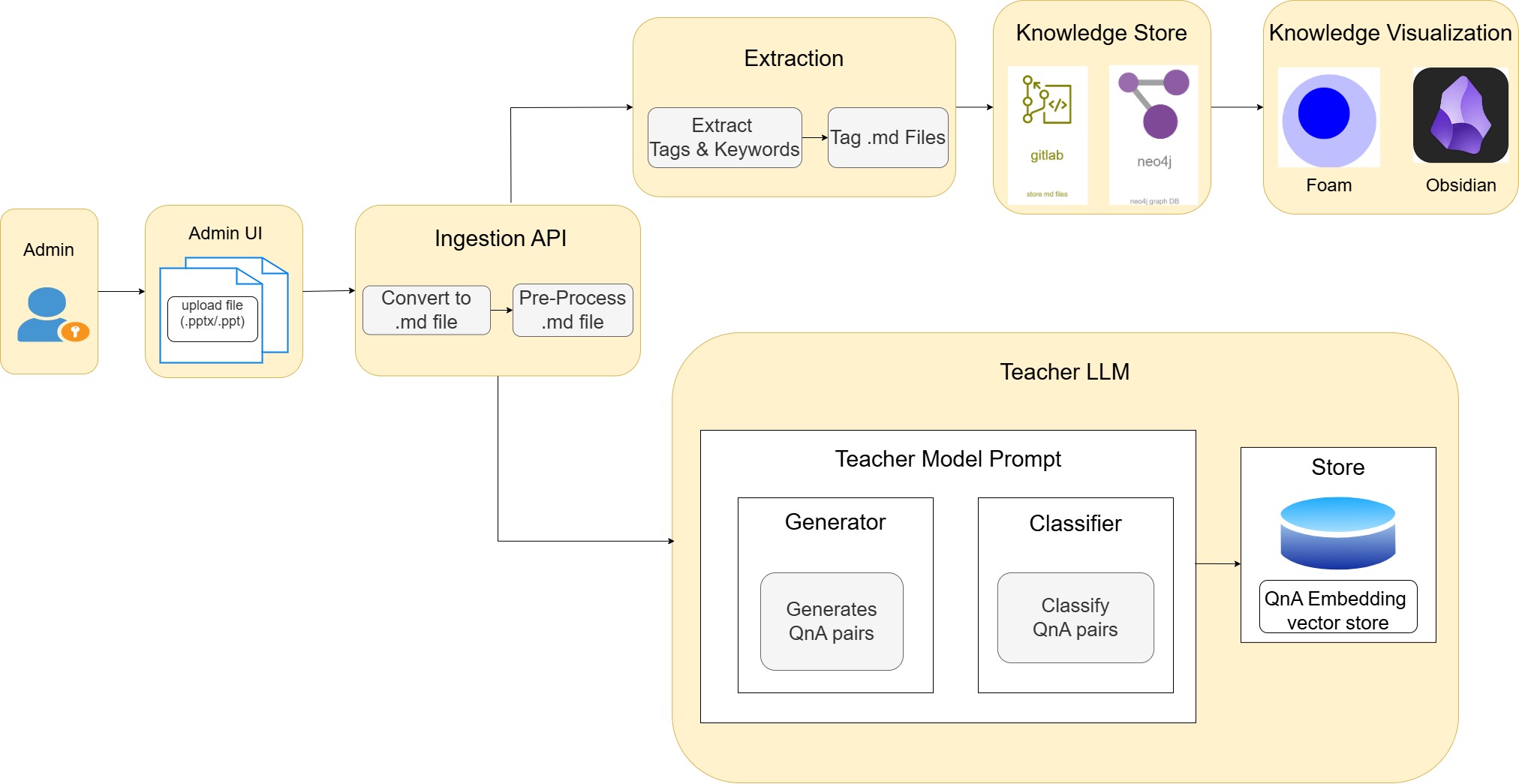}
     \caption{ZeroG - Document Ingestion and Distilled Dataset Generation Pipeline}
   \end{minipage}\hfill
\end{figure}

In addition to the knowledge store, an advanced backend service is established to enhance document processing and retrieval using a Retrieval Augmented Generation (RAG) approach. This service semantically chunks documents and stores these chunks within a vector store collection. By generating vector embeddings for each document using simple embedding models like the \emph{all-MiniLM-L6-v2} \cite{huggingface_minilm}, we can effectively perform similarity search of user queries with relevant document information.

To efficiently store these embeddings and ensure the service is production-ready, we use \emph{pgvector} \cite{kane2023pgvector}, a PostgreSQL extension designed for managing high-dimensional vectors. Pgvector’s scalability and performance make it ideal for handling large volumes of document embeddings, ensuring optimal retrieval speeds during queries.

As seen in \emph{Figure 1}, the service incorporates a question generator module that utilizes a LLM with higher parameters, such as \emph{Mixtral-8x7b} \cite{huggingface_mixtral}, which is more powerful and capable of handling more complex queries, and we denote this as the Teacher Model. Initially, we used \emph{Mixtral-8x7b} but later moved to \emph{Qwen2-7b} \cite{qwen2} based on experiments showing improved QnA pairs. This model, based on few-shot prompting, generates QnA pairs from document contents and classifies them into certain broad groups as labels included in the prompt, embedding the generated questions alongside the plain text in the QnA vector store collection. We also capture the document metadata used to generate the QnA pairs, such as the document title, author, and date, and store it along with the generated question label as part of the QnA metadata. The knowledge generated by the teacher model is then distilled \cite{hinton2015distillingknowledgeneuralnetwork} for the student model to operate upon, enhancing the system's ability to provide precise and contextually relevant responses to user queries, further optimizing the document retrieval process.

\subsection{Retrieval and Response Generation - Knowledge Distillation}
We perform an MMR (Maximal Marginal Relevance) \cite{adams2022combiningstateoftheartmodelsmaximal} search on the incoming user query against the question embeddings stored in the QnA vector. Initially, we used cosine similarity for query retrieval, but transitioned to MMR due to its superior ability to balance relevance with novelty. MMR reranks results by considering both the relevance to the query and the novelty compared to other selected documents, ensuring that the retrieved set is both relevant and diverse.

\begin{figure}[h]
   \begin{minipage}{\textwidth}
     \centering
     \includegraphics[height=8.5cm]{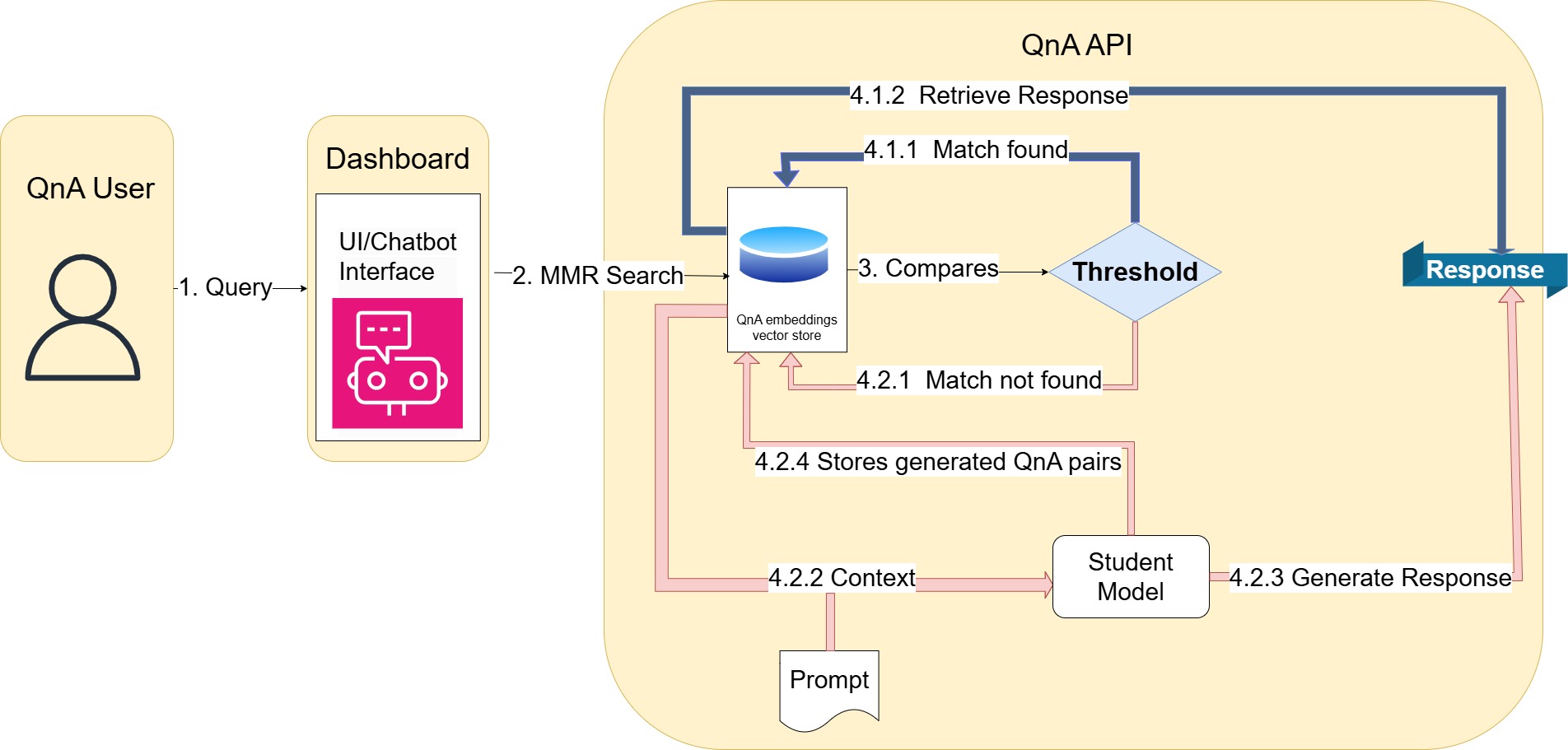}
     \caption{ZeroG - Response Retrieval Pipeline}
   \end{minipage}\hfill
\end{figure}

When a query's MMR score surpasses the established threshold of \emph{0.93}, the response of the corresponding QnA pair is returned to the user. This threshold was determined through extensive experimentation to balance precision and recall, ensuring high accuracy of responses. As depicted in \emph{Figure 2}, if the score falls below this threshold, the top three most relevant QnA pairs, identified through iterative testing and validation, are sent to a smaller student model. For our solution, we used \emph{Phi-3-mini} \cite{huggingface_phi3}, a resource-efficient model. The student model used a zero-shot prompt to effectively utilize the provided context of QnA pairs and generate an accurate response to the user's query. Both the query and the generated response are then added to the existing QnA pairs in the vector database, continuously enhancing the system's knowledge base.

\section{Discussion and Analysis}
\subsection{Similarity Search}
Transitioning from cosine similarity to MMR in our retrieval process has markedly enhanced the effectiveness and accuracy of our QnA system. While cosine similarity is adept at measuring direct vector similarities, it often fails to introduce diversity in the results, leading to redundant or overly similar document selections. This limitation can result in missing out on valuable, diverse insights that are crucial for comprehensive understanding.

In contrast, MMR evaluates both the relevance of a document to the query and its novelty compared to others. This dual consideration ensures a balanced and comprehensive set of results. Our implementation showed an increase in retrieval accuracy, particularly notable when the data quality is not optimal. In our experiments, where the data was not extensively pre-processed, we observed up to a 12\% increase in accuracy compared to using only cosine similarity. However, when data quality is higher and richer, both cosine similarity and MMR tend to perform similarly.

MMR's reranking \cite{pickett2024betterragusingrelevant} capability reduces redundancy in responses, ensuring users receive a wider range of information. This diversity enhances answer quality and broadens user perspectives, which is especially valuable in complex queries that benefit from multiple viewpoints. Adopting MMR has enabled our system to deliver more accurate and contextually rich responses, significantly boosting user satisfaction and system reliability. These improvements highlight the vital role of selecting appropriate retrieval methods to optimize the performance of knowledge extraction systems.

\subsection{Optimizing Model Performance through Effective Prompt Tuning}
In our approach, effective prompt tuning was pivotal in optimizing both the teacher and student models. The teacher model required extensive tuning through few-shot examples and carefully crafted prompts. This process guided the model to understand and extract relevant information from complex documents, enabling it to generate comprehensive QnA pairs.

For the student model, the focus was on zero-shot prompting to utilize the context provided by the QnA pairs from the teacher model. This required designing prompts that directed the student model accurately without overwhelming it, allowing efficient processing of selected QnA chunks. This method ensured that responses were both accurate and contextually relevant.

The interplay between model, data, and prompt tuning is crucial. Model tuning ensures optimal performance, data tuning refines input accuracy, and prompt tuning directs the model's focus. Together, these elements create a synergistic effect that enhances the system's performance. This holistic approach highlights the importance of integrating model, data, and prompt tuning to achieve the best outcomes in knowledge extraction and response generation, ensuring models deliver precise and meaningful responses. Balancing these elements maximizes their potential and effectiveness.

\subsection{Knowledge Distillation - Optimizing Contextual Integrity}
\begin{figure}[h]
   \begin{minipage}{\textwidth}
     \centering
     \includegraphics[height=13cm]{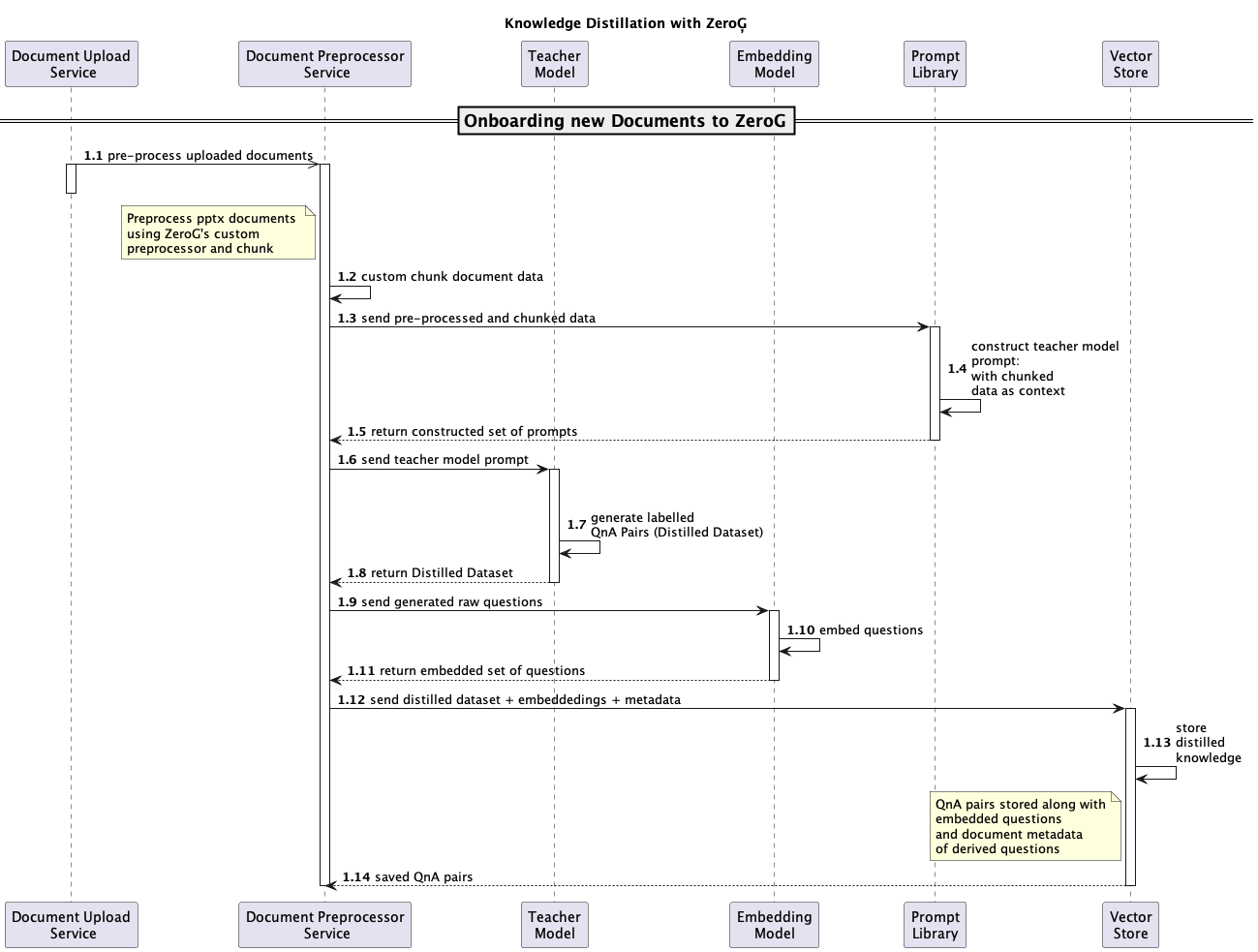}
     \caption{ZeroG - Knowledge Distillation: Generating Distilled Data}
   \end{minipage}\hfill
\end{figure}
Knowledge distillation \cite{xu2024surveyknowledgedistillationlarge} represents a pivotal advancement in the ZeroG framework, offering a significant improvement over traditional RAG solutions. While scaling models to larger sizes can enhance performance, it often becomes computationally prohibitive for widespread deployment. In contrast, knowledge distillation provides an efficient alternative, particularly in mitigating issues such as hallucinations and maintaining contextual relevance.

In traditional RAG-based solutions, there is a higher likelihood of models hallucinating or producing out-of-context responses. This is partly due to the expansive nature of LLMs attempting to generate responses without strict contextual constraints. Knowledge distillation mitigates this issue by employing a black-box approach \cite{li-etal-2023-symbolic, shridhar2023distillingreasoningcapabilitiessmaller}, where a larger teacher model generates a distillation dataset, as depicted in \emph{Figure 3}. This dataset consists of labeled QnA pairs created from processed document data and serves as a foundational element for augmenting the responses of a smaller student model \cite{chen2024rolesmallmodelsllm} without relying on intermediate features or logits,

The distillation dataset, formed through the teacher's generation of pseudo labels, enables the student model to replicate the contextual behavior of the teacher model. By utilizing this dataset, we can ensure the student model's responses are well-aligned with the intended context, thereby enhancing output quality and significantly reducing the risk of hallucinations, while keeping compute overheads to a minimum. By focusing strictly on the provided context, the student model effectively receives only relevant information in a dense (high correlation) embedding, ensuring the model remains within the desired context (groundedness), further minimizing the chance of producing out-of-context answers (hallucinations).

\begin{figure}[h]
   \begin{minipage}{\textwidth}
     \centering
     \includegraphics[height=16cm]{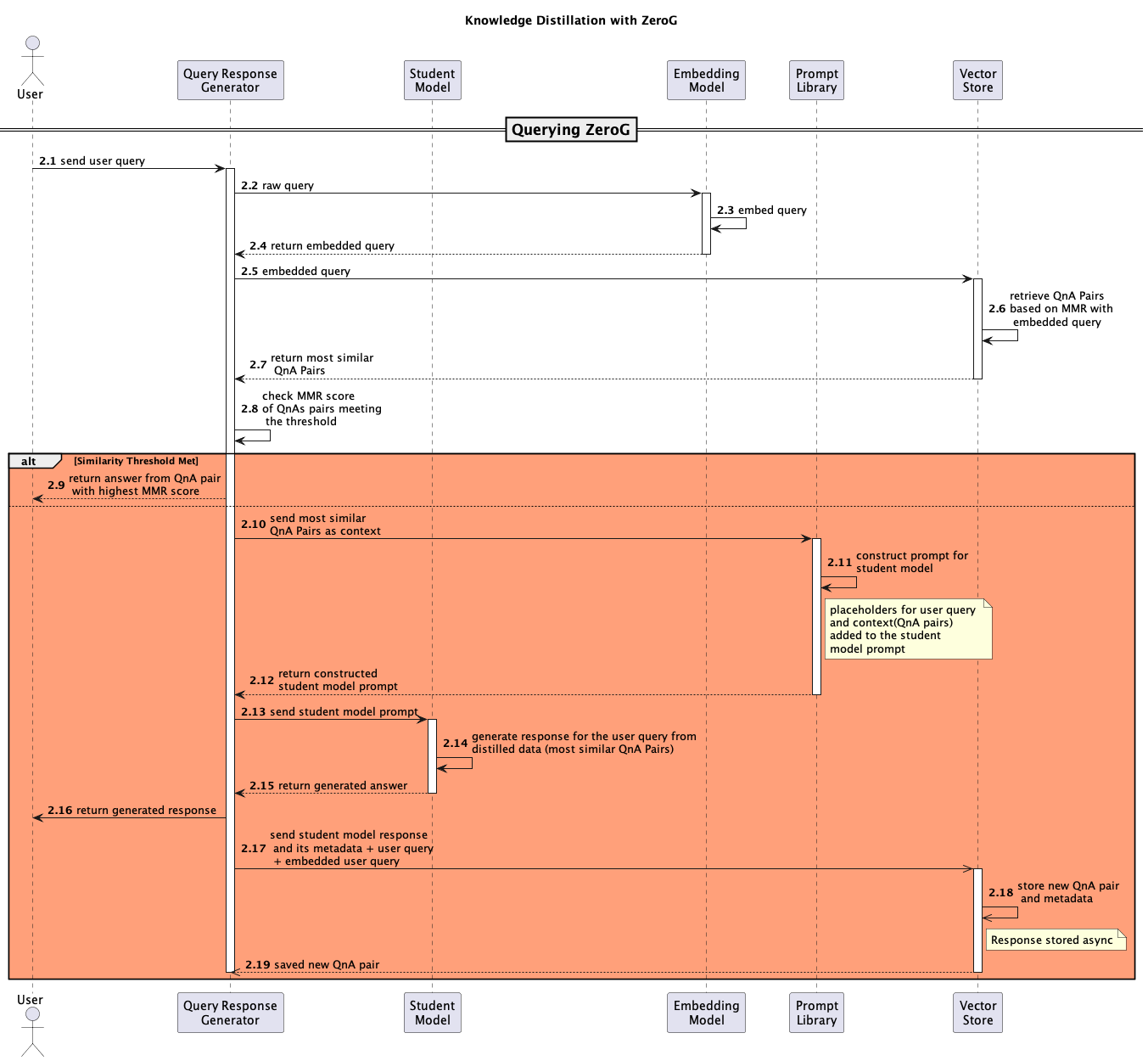}
     \caption{ZeroG - Knowledge Distillation: RAG pipeline}
   \end{minipage}\hfill
\end{figure}

The use of smaller models is particularly advantageous in environments with limited computational resources \cite{hsieh2023distillingstepbystepoutperforminglarger}. Knowledge distillation brings together highly correlated, in-context factoids from source documents through teacher model to smaller models, allowing any small language model (SLM) to achieve higher than rated performance while significantly reducing their CPU and memory footprint. This makes them ideal for computationally intensive tasks such as information retrieval when dealing with a large corpus of documents.

The smaller language model in ZeroG is designed to provide faster responses with acceptable accuracy for moderately complex queries. It 'learns' to utilize the context of responses generated by the teacher model, striking a balance between response time and model capability. This ensures an optimal user experience by maintaining accuracy while reducing resource demands. 

As illustrated in \emph{Figure 4}, when the MMR score threshold is not met, the process involves constructing a prompt for the student model that incorporates placeholders for the user query and context based on retrieved QnA pairs. This prompt enables the student model to effectively leverage the distilled knowledge from the teacher model. The student model then generates a response using this distilled information, ensuring that it can provide relevant answers. The response from the student model is subsequently returned to the user, and the new QnA pair is stored in the Vector Store for future reference. This approach highlights the effectiveness of knowledge distillation in enhancing the performance and responsiveness of smaller models in the ZeroG framework.

Our approach aims to enhance the zero-shot performance of smaller LLMs by providing them with a context of instruction-like prompt-response pairs. This ensures that the model deals only with in-context data, effectively normalizing its responses. In summary, knowledge distillation in ZeroG effectively alleviates the challenges associated with large-scale models, offering a solution that is both efficient and contextually accurate. By leveraging this technique, ZeroG achieves a balance between computational efficiency and model performance, ensuring a reliable and user-focused experience.

\pagebreak

\section{Comparative Study}
We conducted a comparative analysis between ZeroG, which utilizes knowledge distillation, and traditional RAG solutions. The key differences highlight the advantages of employing knowledge distillation in terms of accuracy, response time, reducing hallucinations, and overall performance.

ZeroG leverages knowledge distillation to significantly enhance the model's accuracy, reduce latency, and mitigate hallucinations compared to RAG-based solutions. While RAG solutions are effective, they often suffer from longer response times and a higher likelihood of producing out-of-context or hallucinatory responses. In contrast, ZeroG's approach ensures responses are more contextually relevant and timely.

Our evaluation was performed using the DeepEval \cite{confident2024deepeval} module, an LLM evaluation framework. This module allows for comprehensive assessment of various metrics including accuracy, comprehensibility, response time, and safety and bias incidents. 

The following table (\emph{Table 1}) presents the performance of ZeroG with knowledge distillation compared to a RAG-based solution without knowledge distillation, across several key metrics:

\begin{table}[h!]
    \centering
    \resizebox{\textwidth}{!}{%
    \begin{tabular}{lcc}
        \toprule
        \textbf{Metric} & \textbf{Without Knowledge Distillation (RAG)} & \textbf{With ZeroG (Knowledge Distillation)} \\
        \midrule
        Accuracy & 73\% & 87.5\% \\
        Comprehensibility & 91\% & 97\% \\
        Response Time (Latency) & 17 sec & < 6 sec \\
        \# of Safety \& Bias Incidents & $\approx 1 \text{ in a 1000}$ & $\approx 1 \text{ in a 1000}$ \\
        \bottomrule
    \end{tabular}%
    }
    \caption{Comparison of ZeroG with Knowledge Distillation versus RAG-Based Approach}
    \label{tab:comparison}
\end{table}

The use of knowledge distillation fundamentally differentiates ZeroG from traditional RAG approaches, allowing for the transfer of nuanced contextual understanding from a larger teacher model to a smaller student model. This transfer results in responses that are not only more accurate but also more contextually aligned with user queries, thereby reducing the likelihood of out-of-context responses commonly observed in RAG solutions.

In addition, ZeroG demonstrates improved efficiency through reduced response times, which is crucial for user satisfaction in real-time applications. Knowledge distillation enhances this efficiency by streamlining the model's ability to generate responses based on distilled knowledge rather than solely relying on the expansive retrieval processes typical of RAG.

ZeroG shows no adverse impact on safety, ensuring the model behaves consistently within its expected behavior patterns and aligns with the intended questions and context provided. When the model is uncertain about an answer, it appropriately responds that it doesn't know, rather than generating potentially misleading information. This is achieved through the use of prompt-level guardrails \cite{ajwani2024plugplaypromptsprompt}, which steer the model's responses in the direction of the context. By implementing these guardrails, we aim to steer the model away from generating biased and toxic outputs, ensuring that the responses are relevant, accurate, and safe. To further mitigate any safety concerns, techniques like NeMo Guardrails \cite{rebedea-etal-2023-nemo} and fine-tuning can be employed \cite{kang2023knowledgeaugmentedreasoningdistillationsmall} to eliminate risks, ensuring a reliable and safe user experience.

Overall, these metrics collectively demonstrate the effectiveness of knowledge distillation in enhancing the accuracy, speed, and clarity of responses, making ZeroG a more effective solution than traditional RAG methods.

\section{Conclusion}
In conclusion, ZeroG presents a transformative approach to managing and extracting knowledge from digital documents, addressing the inherent challenges of handling complex data in real-time. Through the integration of knowledge distillation and prompt tuning, ZeroG enhances the performance of smaller language models, significantly improving response accuracy and reducing latency. This innovative approach mitigates the risks of hallucinations commonly associated with large language models in a statistically significant manner and provides a scalable solution for modern document management.

By leveraging advanced techniques like semantic management, metadata utilization, and graph databases, ZeroG ensures precise, contextually relevant responses, thereby streamlining information retrieval processes. The shift from traditional RAG solutions to a knowledge-distilled framework has demonstrated marked improvements in accuracy and efficiency, as evidenced by our comparative analysis.

The adaptability of ZeroG to evolving data needs and its focus on data privacy and computational efficiency make it an invaluable tool for organizations handling sensitive and complex documents. As the demand for efficient document management systems continues to grow, ZeroG stands poised as a robust solution, enhancing productivity and user experience.

\section{Future Areas of Research}
Future research will explore the expansion of ZeroG's capabilities, including the integration of additional document types and further optimization of its retrieval processes. This ongoing development promises to further solidify ZeroG's position as a leading solution in the landscape of digital document management.

\subsection{Furthering Distillation}
Explore \textbf{white-box distillation}, which involves leveraging the internal states of the teacher model to provide transparency in the student model's training process. By utilizing output distributions and intermediate features from the teacher LLMs—collectively known as feature knowledge—we can develop cost-effective yet powerful models such as DistilBERT \cite{sanh2020distilbertdistilledversionbert} and QuantizedGPT \cite{yao2022zeroquantefficientaffordableposttraining}.

\subsection{Improve Context Quality}
Broaden and improve the knowledge transfer process by integrating feedback on the student model's performance and utilizing feature knowledge for greater benefits. While much of the work in LLM knowledge distillation has concentrated on skill transfer, there is a promising opportunity to enhance attributes like reliability, honesty, and safety. Emphasizing these qualities will contribute to building trustworthy and efficient models, fostering further progress in this area.

\subsection{Improving Embedding Space \& Vector Shifting}
Minimize the overhead (\textbf{token reduction}) associated with token generation while preserving output quality. These mechanisms reduce the number of generated tokens and enhance the efficiency of ZeroG. This can be accomplished by creating more complex (longer) tokens that span full domain-specific terms or more, thereby increasing the density of the network and reducing computational overhead.

\subsection{Dual Model Strategy}
Employ a \textbf{dual-model strategy} where one model analyzes user questions and categorizes them into topics, while a larger model is fed contextual information. Instead of producing lengthy and complex responses, this larger model will generate concise topics \cite{liang2024languashrinkreducingtokenoverhead} that various SLMs can then use to formulate specific responses. This strategy not only enables the addition of domain-specific tokens but also facilitates workload distribution among different models, potentially sharing a latent space to increase coherence between their outputs \cite{he2023multimodallatentspacelearning}.

\subsection{Pre-computing Latent Space}
This can significantly improve performance by allowing faster retrieval of relevant embeddings, thus minimizing the time spent on generating tokens. A \textbf{shared latent space} will enhance the consistency of generated responses, ensuring that both models align more closely in their outputs. Additionally, by leveraging this shared latent space, we can adapt outputs \cite{Belanec_2023} to address multiple response types at once, effectively catering to different information needs while minimizing token generation costs. An embedding function will link text with generated embeddings and topics, effectively reducing the token space and overhead. By combining the strengths of large models, small models, and extremely small models, we aim to create a more efficient and cost-effective framework for knowledge management and retrieval, thereby enhancing ZeroG's capabilities in handling varied depths of information across diverse document types for user queries.

\section{Acronyms}
\begin{acronym}
  \acro{LLM}{ = Large Language Model} 
  \acro{MMR}{ = Maximal Marginal Relevance}
  \acro{RAG}{ = Retrieval-Augmented Generation}
  \acro{RFP}{ = Request for Proposal}
  \acro{SLM}{ = Small Language Model}
\end{acronym}

\pagebreak

\bibliographystyle{unsrt}
\bibliography{references}

\end{document}